# Judgments of research co-created by generative AI: experimental evidence

Paweł Niszczota[1,*], Paul Conway[2]

[1] Poznań University of Economics and Business, Institute of International Business and Economics, Humans & AI Laboratory (HAI Lab), Poznań, Poland

[2] University of Southampton, School of Psychology, Centre for Research on Self and Identity, Southampton, United Kingdom

* Corresponding author: Paweł Niszczota, Poznań University of Economics and Business, al. Niepodległości 10, 61-875 Poznań, Poland, pawel.niszczota@ue.poznan.pl

**Abstract**

The introduction of ChatGPT has fuelled a public debate on the use of generative AI (large language models; LLMs), including its use by researchers. In the current work, we test whether delegating parts of the research process to LLMs leads people to distrust and devalue researchers and scientific output. Participants (N=402) considered a researcher who delegates elements of the research process to a PhD student or LLM, and rated (1) moral acceptability, (2) trust in the scientist to oversee future projects, and (3) the accuracy and quality of the output. People judged delegating to an LLM as less acceptable than delegating to a human ($d$ = -0.78). Delegation to an LLM also decreased trust to oversee future research projects ($d$ = -0.80), and people thought the results would be less accurate and of lower quality ($d$ = -0.85). We discuss how this devaluation might transfer into the underreporting of generative AI use.

**Keywords:** trust in science, metascience, ChatGPT, GPT, large language models, generative AI, experiment

**JEL codes:** G00, J24, O33, O34

**Acknowledgments**

This research was supported by grant 2021/42/E/HS4/00289 from the National Science Centre, Poland.

**Introduction**

The introduction of ChatGPT appears to have become a tipping point for large language models. It is expected that large language models – such as those released by OpenAI (ChatGPT, GPT-4; OpenAI, 2023, 2022), but also major technology firms such as Google and Meta – will impact the work of many white-collar professions (Alper and Yilmaz, 2020; Eloundou et al., 2023). This includes top academic journals such as *Nature* and *Science* that have already acknowledged the impact it has on the scientific profession, and have started setting some guides on how to use large language models (Thorp, 2023; "Tools such as ChatGPT threaten transparent science; here are our ground rules for their use," 2023). For example, listing ChatGPT as a co-author was deemed inappropriate (Stokel-Walker, 2023; Thorp, 2023). However, the use of such models is not explicitly forbidden – rather, it is suggested that researchers report in which part of the research process did they receive assistance from ChatGPT.

Yet, important questions remain regarding the perceptions of scientists employing in their work (but see, e.g., Dwivedi et al., 2023). Do people view the use of large language models as diminishing the importance, value, and worth of scientific efforts, and if so, which elements of the scientific process does large language model usage most impact? We examine these questions with a study on the perceptions of scientists who rely on a large language model for various aspects of the scientific process.

We anticipated that, overall, people would view delegating aspects of the research process to a large language model as morally worse than delegating to a human, and that doing so would reduce trust in the delegating scientist. Moreover, insofar as people view creativity as a core human trait, especially in comparison to AI (Cha et al., 2020), and some aspects of the research process may entail more creativity than others—such as idea generation and prior literature synthesis (e.g., King, 2023), compared to data identification and preparation, testing framework determination and implementation, or results analysis—we tested the exploratory prediction that the effect of delegation to AI versus a human on moral ratings and trust might be different for these aspects.

We contribute to an emerging literature exploring how large language models can assist research on economics and financial economics. The reader can find a valuable discussion on the use of large language models in economic research in Korinek (2023). A noteworthy empirical study can be found in Dowling and Lucey (2023), who asked financial academics to rate research ideas on cryptocurrency and judged that the output is of fair quality.

## 1. Research questions

We ask two research questions concerning laypeople's perception of the use of large language models in science. First, we tested the hypothesis that people will perceive research assistance from large language models less favourably than the very same assistance from a junior human researcher. In both cases, we assume that the assistance is minor enough to not warrant co-authorship. This levels the playing field for human and AI assistance, as prominent journals have already expressed that large language models cannot be listed as co-authors (Thorp, 2023), as was already done in some papers (e.g., Kung et al., 2022).

Second, we examined in which aspects of the research process are the prospective human-AI disparities the strongest. If – as we hypothesize – delegating to AI is perceived less favourably, then one can assume that delegating such processes to AI will have the greatest potential to devalue work done by scientists.

### 1.1. Participants

To assess the consequences of delegating research processes to large language models, we recruited 441 participants from Prolific (Palan and Schitter, 2018), that had a 98% or higher approval rating, were located and born in the United States, and whose first language was English. As preregistered, 39 participants that didn't correctly answer both attention check questions were excluded, leaving a final sample size of 402 (48.3% female, 49.8% male, and 1.9% selected non-binary or did not disclose). The mean age of participants was 42.0 years ($SD = 13.9$). 97.5% have heard about ChatGPT, and 38.1% interacted with it.

The study was pre-registered at https://aspredicted.org/GVL_MR5. Data and materials are available at https://osf.io/fsavc/?view_only=a49d1e7ad73446df8bdaede2024e2b6d.

### 1.2. Experimental design

We conducted a mixed-design experiment. We randomly allocated participants to one of two conditions between-subjects. Participants rated a distinguished senior researcher who delegated a part of the research process to either another person—specifically, a PhD student with two years experience in the area (human condition), or to a large language model such as ChatGPT (large language model condition). Each participant rated the effect of such delegation on each of the five parts of the research process discussed in Cargill and O'Connor (2021): idea generation, prior literature synthesis, data identification and preparation, testing framework determination and implementation, and results analysis. Notably, Dowling and Lucey (2023) used all of these except results analysis to assess the quality of ChatGPT's output. We rephrased the two last research processes for clarity.

For each research process, participants rated three items:

1. I think that it is morally acceptable for a scientist to delegate - in such a scenario - the following part of the research process (after giving credit in the Acknowledgments);
2. I think that a scientist that delegated the part of the research process shown below should be trusted to oversee future research projects;
3. I think that delegating this part of the research process will produce correct output and stand up to scientific scrutiny (e.g., results would be robust, reliable, and correctly interpreted).

We expected the first two items to correlate with one another, but not necessarily with the third. While people might acknowledge that AI might be better than humans in some tasks, they often exhibit an aversion toward the use of algorithms (Dietvorst et al., 2015).

## 2. Results
## 2.1. Preliminary analysis

Prior to presenting regression results, we examined how answers correlated with each other. As expected, moral acceptability ratings[1] correlated highly with trust to oversee future projects, $r = 0.81$, $p < 0.001$. However, moral acceptability ratings also correlated highly with accuracy ratings, $r = 0.81$, $p < 0.001$. Similarly, trust ratings correlated highly with accuracy, $r = 0.80$, $p < 0.001$.

However, it remains possible that the relationship between such perceptions was lower when the scientist delegated to a large language model instead of a human. To determine this, we conducted a regression analysis treating one item as the dependent variable, and another as the independent variable, but we added an interaction with a dummy variable across delegation condition. Results, presented in **Table 1**, suggest that the strength of the relationship between moral acceptability, trust, and accuracy either becomes stronger when delegating to a large language model (rather than a human) or is not statistically different. Therefore, people evaluated moral acceptability, trust, and accuracy in a similar manner in each condition.

**Table 1. The interrelationship between ratings of three items (moral acceptability, trust to oversee, and accuracy)**

|  | Moral acceptability | Trust | Accuracy |
|---|---|---|---|
| (Intercept) | 0.12 * (0.05) | 0.11 * (0.05) | 0.10 (0.05) |
| Trust | 0.68 *** (0.06) |  |  |
| Large language model (1 = yes, 0 = no) | -0.17 ** (0.06) | -0.15 * (0.06) | -0.16 * (0.07) |
| Trust × LLM | 0.14 * (0.07) |  |  |
| Accuracy |  | 0.67 *** (0.06) | 0.72 *** (0.06) |

---

[1] The correlations were based on mean ratings from the five research processes.

|  | Moral acceptability | Trust | Accuracy |
|---|---|---|---|
| Accuracy × LLM |  | 0.16 * (0.07) | 0.07 (0.07) |
| N | 402 | 402 | 402 |
| $R^2$ adjusted | 0.660 | 0.656 | 0.636 |

*Notes:* Ratings are means for five research processes. Moral acceptability, trust, and accuracy scores are standardized to facilitate the interpretability of the coefficient for LLM (which corresponds to the effect of delegating to the LLM (relative to the human) when trust or accuracy is at its mean level).

\* $p < 0.05$  \*\* $p < 0.01$  \*\*\* $p < 0.001$

### 2.2. Pre-registered analysis

We present the results of the pre-registered analysis in **Table 2**. Consistent with the hypothesis, people rated delegating the research process to a large language model as less morally acceptable and reported lower trust towards this scientist to oversee future research projects. Moreover, people also rated delegating to an LLM as producing less correct output. The effect of delegating to a large language model (relative to delegating the same to a PhD student) was similar for all three items and thus results from the combined dataset ("All items and processes") can serve as a benchmark for future studies.

For readers accustomed to Cohen's *d* (Cohen, 1988), the effect sizes (and 95% confidence intervals) of delegating to a large language model instead of a human were large: $d = -0.78$ [-0.99, -0.58] for moral acceptability, $d = -0.80$ [-1.00, -0.60] for trust, and $d = -0.85$ [-1.06, -0.65] for accuracy.

### 2.3. Exploratory analysis

Table 2 and Figure 1 present how ratings varied across the five research processes and conditions. The adverse effect of delegating to a large language model was strongest for the "Testing and interpreting the theoretical framework" process and weakest for the "Statistical result analysis" process. However, the patterns were robust for each of the five research elements, in the $d = -0.81$ (large effect) to -0.51 (medium effect) range. Therefore, despite some variation across research processes, people nonetheless judged delegation of any process to an LLM as worse than to a human.

**Table 2. Perceptions of delegating parts of the research process**

| | All items and processes | Items | | | Research processes | | | | |
|---|---|---|---|---|---|---|---|---|---|
| | | Moral acceptability | Trust | Correctness | Idea generation | Prior literature synthesis | Data identification and preparation | Testing and interpreting the theoretical framework | Statistical result analysis |
| (Intercept) | 5.34 *** (0.09) | 4.92 *** (0.46) | 4.90 *** (0.50) | 5.16 *** (0.46) | 4.83 *** (0.53) | 5.24 *** (0.48) | 5.34 *** (0.48) | 5.52 *** (0.53) | 4.67 *** (0.51) |
| Large language model | -1.07 *** (0.12) | -1.01 *** (0.13) | -1.13 *** (0.14) | -1.09 *** (0.13) | -0.89 *** (0.15) | -1.12 *** (0.14) | -1.23 *** (0.13) | -1.30 *** (0.15) | -0.85 *** (0.14) |
| Item = *Correctness* | -0.16 *** (0.03) | | | | -0.08 (0.06) | -0.16 ** (0.06) | -0.25 *** (0.06) | -0.07 (0.06) | -0.22 *** (0.06) |
| Item = *Trust* | 0.07 * (0.03) | | | | 0.11 (0.06) | 0.04 (0.06) | -0.05 (0.06) | 0.20 *** (0.06) | 0.02 (0.06) |
| Research process = *Prior literature synthesis* | 0.21 *** (0.04) | 0.26 *** (0.08) | 0.19 ** (0.06) | 0.18 ** (0.07) | | | | | |
| Research process = *Data identification and preparation* | 0.28 *** (0.04) | 0.39 *** (0.08) | 0.23 *** (0.06) | 0.22 ** (0.07) | | | | | |
| Research process = *Testing and interpreting the theoretical framework* | -0.11 ** (0.04) | -0.15 (0.08) | -0.05 (0.06) | -0.14 * (0.07) | | | | | |
| Research process = *Statistical result analysis* | 0.11 * (0.04) | 0.18 * (0.08) | 0.09 (0.06) | 0.05 (0.07) | | | | | |
| Age | | 0.01 (0.00) | 0.01 (0.01) | 0.01 (0.00) | 0.00 (0.01) | 0.01 (0.00) | 0.01 * (0.00) | 0.00 (0.01) | 0.02 ** (0.01) |

|  | All items and processes | Items | | | Research processes | | | | |
| --- | --- | --- | --- | --- | --- | --- | --- | --- | --- |
|  |  | Moral acceptability | Trust | Correctness | Idea generation | Prior literature synthesis | Data identification and preparation | Testing and interpreting the theoretical framework | Statistical result analysis |
| Heard of ChatGPT |  | 0.01 (0.42) | 0.14 (0.45) | -0.23 (0.41) | 0.26 (0.48) | 0.09 (0.44) | -0.10 (0.43) | -0.38 (0.48) | 0.00 (0.47) |
| Interacted with ChatGPT |  | 0.02 (0.13) | 0.06 (0.14) | -0.12 (0.13) | 0.26 (0.16) | -0.00 (0.14) | -0.01 (0.14) | -0.12 (0.16) | -0.19 (0.15) |
| Gender | Yes | Yes | Yes | Yes | Yes | Yes | Yes | Yes | Yes |
| **Random Effects** | | | | | | | | | |
| $\sigma^2$ | 1.12 | 1.20 | 0.77 | 0.93 | 0.79 | 0.64 | 0.64 | 0.70 | 0.71 |
| $\tau_{00}$ | 1.40 id | 1.41 id | 1.78 id | 1.45 id | 1.96 id | 1.62 id | 1.58 id | 1.99 id | 1.84 id |
| ICC | 0.56 | 0.54 | 0.70 | 0.61 | 0.71 | 0.72 | 0.71 | 0.74 | 0.72 |
| $N_{participants}$ | 402 id | 402 id | 402 id | 402 id | 402 id | 402 id | 402 id | 402 id | 402 id |
| N | 6030 | 2010 | 2010 | 2010 | 1206 | 1206 | 1206 | 1206 | 1206 |
| Marginal $R^2$ / Conditional $R^2$ | 0.111 / 0.606 | 0.106 / 0.589 | 0.119 / 0.733 | 0.127 / 0.660 | 0.078 / 0.736 | 0.131 / 0.752 | 0.162 / 0.757 | 0.151 / 0.779 | 0.096 / 0.748 |

*Notes:* Linear-mixed models were estimated using *R* packages *lme4* (Bates et al., 2015) and *lmerTest* (Kuznetsova et al., 2017). The baseline values are "Moral acceptability" for Item, "Idea generation" for Research process, and female for gender. All variables bar age and gender are dummy variables, taking the value 1 if the variable is equal to what the variable's name implies, and 0 otherwise. * $p < 0.05$  ** $p < 0.01$  *** $p < 0.001$

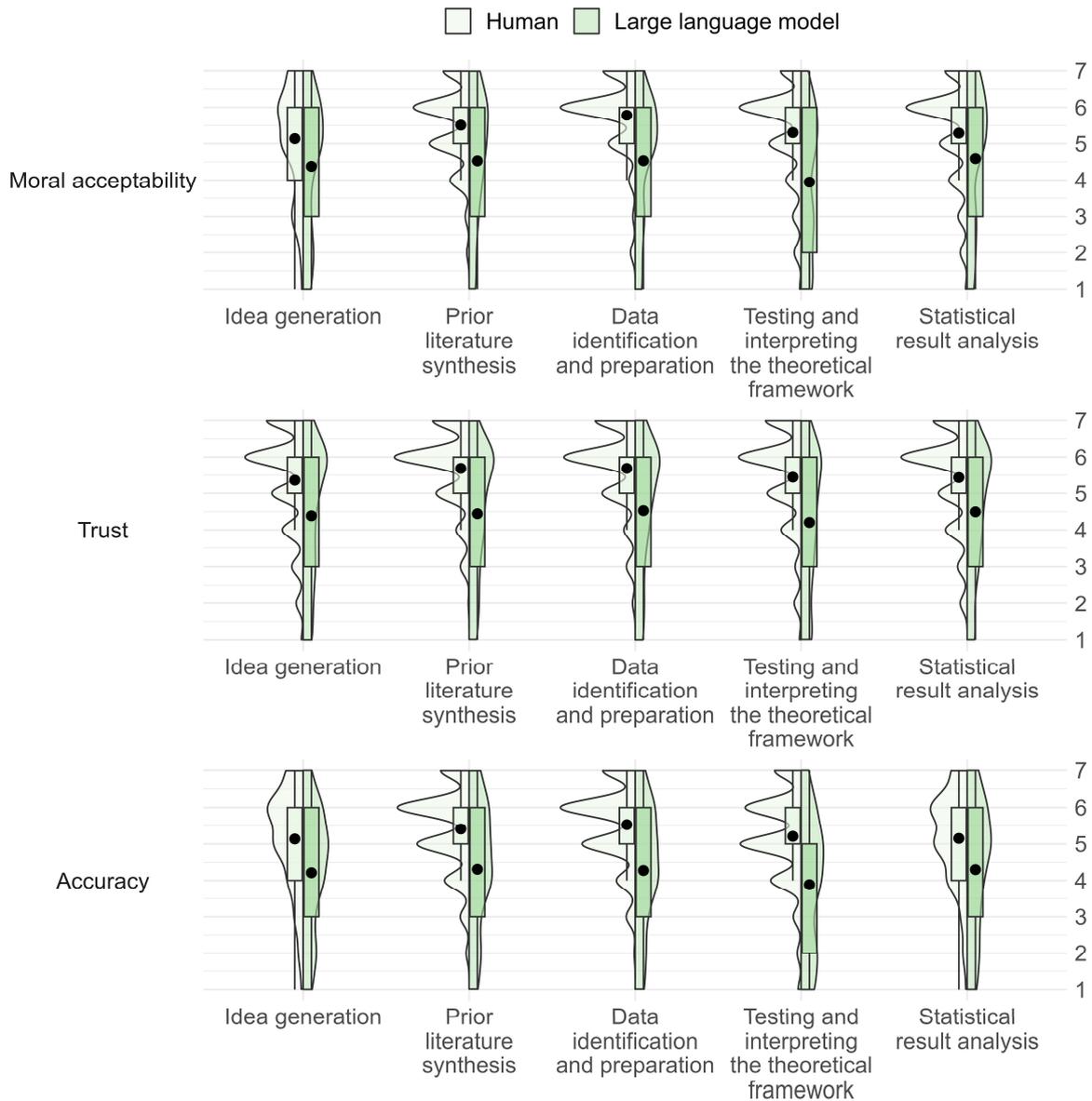

**Figure 1. Ratings of moral acceptability, trustworthiness of scientist to oversee future projects after delegation, and accuracy of science across research processes**

*Notes:* Dots represent means.

## 3. Discussion

Overall, these results suggest that people have clear, strong negative views of scientists delegating any aspect of the research process to ChatGPT or similar large language models compared to a PhD student. They rated delegation to an LLM as less morally acceptable, a scientist choosing such delegation as less trustworthy for future projects, and they rated the output of such delegation as less accurate and of lower quality. These ratings held across all five aspects of the research process identified in past work: idea generation, prior literature synthesis, data identification and preparation, testing framework determination and implementation, and results analysis (Cargill and O'Connor, 2021). Although people showed the strongest differentiation between LLMs and human researchers for testing and interpreting the theoretical framework, and the weakest for statistical result analysis, the effect size for all five was substantial, with Cohen's *d*'s that would be conventionally described as medium to large (Cohen, 1988), but can, based on effect sizes that are observed in psychological research, be considered large to very large (Funder and Ozer, 2019).

Note that, as expected, moral ratings and trust in scientists were highly correlated, but in addition, both correlated highly with perceptions of accuracy and scientific quality. One possibility for this pattern is that people think delegating to LLMs is immoral and untrustworthy precisely because they view the output of such programs as scientifically questionable. This pattern leaves open the possibility that, with further advancement in AI, if the perceived scientific quality of LLMs increases, people may view delegation to such programs as less problematic.

Nonetheless, these results have clear implications for researchers considering use of ChatGPT or other LLMs. At least in their current state, people view such delegation as seriously problematic—as immoral, untrustworthy, and scientifically unsound. This view extends to all aspects of the research process. Therefore, there does not appear to be a widely approved way for researchers to incorporate LLMs into the research process without compromising their work's perceived quality and integrity.

It is worth noting that the current work examined the case where the researcher honestly reports the use of the LLM in the acknowledgments section, as recommended by leading journals such as *Science* (Thorp, 2023). Moreover, the current work examined the case where the researcher delegating to a PhD student—essentially the control condition—features them only in the acknowledgments section rather than as a co-author. Arguably, people may view doing so as ethically questionable as the graduate student would have earned authorship according to common ethical guidelines such as those published by the American Psychological Association (2019). Therefore, the current findings represent a plausible best-case scenario—it is plausible that people would have even stronger negative reactions to a researcher who employed LLMs without revealing their use, essentially taking credit for the output of an algorithm, or compared to a researcher giving their PhD student colleague full authorship credit. These findings underscore the depth of the antipathy toward researchers using LLMs at this time.

### 3.1. Limitations

Like all studies, the current work suffers from some limitations. First, we compared delegation to LLMs to a second-year PhD student—a human with presumably sufficient competence as to normally warrant authorship in scientific publications. Naturally, the choice of comparison target should affect responses. For example, people may think that LLMs will produce more accurate output than, say, a four-year-old or someone who is illiterate. Future work could plausibly test how people perceive LLMs compared to a wide range of targets. However, we elected to begin by testing LLMs against someone who would likely otherwise participate in the scientific process.

Second, we examined the perceptions of laypeople who may have only vague familiarity or understanding of the scientific process. It remains to be seen whether journal editors, reviewers, senior university officials, and others who intimately understand the research process and evaluate scientists share the same views. It may be that with such familiarity, people perceive it more permissible to use LLMs for specific aspects such as data analysis. Findings might also differ using a different split of the research process, perhaps one that includes more fine-grained elements like generating figures based on data computed by humans (Cargill and O'Connor, 2021; Dowling and Lucey, 2023).

Likewise, we examined only perceptions of a scientist operating in a particular area, namely a researcher specializing in economics, finance, and psychology. It remains possible that people hold less-negative views of LLM usage in other branches of science, e.g., perhaps for papers in astrophysics requiring complex calculations. Along the same lines, results may be moderated by the perceived goals of the scientist—for example, it seems likely that people would not hold the same negative impression of research specifically designed to illustrate the uses and limitations of ChatGPT itself (e.g., Kung et al., 2022).

Moreover, we asked people about a hypothetical scientist; it remains possible that asking about a specific (e.g., famous, eminent, trusted) scientist, people demonstrate lower aversion to LLM use—perhaps because they may infer this trusted scientist would only use LLMs if they had specialist knowledge that doing so was worthwhile and not likely to corrupt the research process. In other words, people may moderate inferences about the use of LLMs depending on their prior knowledge and evaluations of a specific scientist.

Finally, the current work examined American participants. It remains possible that results may vary in other populations; for example, Americans tend to view AI more critically than people in China (Wu et al., 2020). Furthermore, not all scientists have equal access to state-of-the-art language models. For example, people from China cannot access these models (Wang, 2023), and Italy has banned access to ChatGPT, at least temporarily (Satariano, 2023). So, perceptions of research delegated to LLM may vary somewhat with access to such models or which models are popular or available.

**Conclusions**

Overall, the current findings suggest that people have strongly negative views of delegating any aspect of the research process to large language models like ChatGPT compared to a junior human scientist: people rated doing so more immoral, more untrustworthy, and the results as less accurate and of lower quality. These findings held for five aspects of the research process from idea generation to data analysis. Therefore, researchers should employ caution when considering whether to incorporate ChatGPT or other large language models into their research. It appears that even when disclosing such practices according to modern standards, doing so may powerfully reduce perceptions of scientific quality and integrity.